\documentclass[11pt]{article}
\usepackage{jheppub}
\usepackage{tikz,subcaption}

\newcommand{\be}{\begin{equation}}
\newcommand{\ee}{\end{equation}}
\newcommand{\bea}{\begin{eqnarray}}
\newcommand{\eea}{\end{eqnarray}}

\newcommand{\cA}{\mathcal{A}}
\newcommand{\cN}{\mathcal{N}}
\newcommand{\cS}{\mathcal{S}}

\title{Antipodal Symmetry of Two-Loop MHV Amplitudes}

\author{Yu-Ting Liu$^{1,2}$}
\affiliation{$^1$ SLAC National Accelerator Laboratory, Stanford University, Stanford, CA 94309, USA}
\affiliation{$^2$ Kavli Institute for Theoretical Physics, UC Santa Barbara, Santa Barbara, CA 93106, USA}
\emailAdd{aytliu@stanford.edu}

\preprint{SLAC-PUB-17696}

\abstract{
I present a conjecture that all two-loop MHV amplitudes in planar $\mathcal{N} = 4$ super-Yang-Mills theory possess an antipodal symmetry when evaluated on parity-even kinematics. The symmetry acts as a change of basis on the symbol letters, followed by the antipode operation associated with the Hopf algebra structure of multiple polylogarithms. At the symbol level, I provide the symmetry map explicitly for amplitudes with up to eight external particles, and also provide evidence at all multiplicities. Intriguingly, the map acts as an isomorphism on the normal fans of the Newton polytopes of the symbol letters. The conjectured symmetry is one of the rare known cases where the antipode map shows up in physically important examples.}

\begin{document}
\maketitle

\section{Introduction}\label{sec:intro}
The $\cN=4$ super-Yang-Mills (SYM) theory is a toy model for QCD, with highly enhanced symmetry as can be seen by the high number of supersymmetry generators. Due to its vanishing beta function, $\cN=4$ super-Yang-Mills theory is conformal even at the quantum level. In addition to the manifest superconformal symmetry, amplitudes in the planar limit possess a hidden \emph{dual} superconformal symmetry \cite{Drummond:2007au,Drummond:2008vq}. This hidden symmetry is related to the amplitude/Wilson loop duality (see ref.~\cite{Alday:2008yw} for a review). The dual conformal symmetry of the amplitude is nothing but the ordinary conformal symmetry of the dual Wilson loop.

Recently, a new duality was discovered between the six-particle MHV amplitude in parity-even kinematics and a three-particle form factor \cite{Dixon:2021tdw}, based on empirical evidence up to seven loops. (The amplitude and the form factor had previously been bootstrapped to seven and eight loops respectively \cite{Caron-Huot:2019vjl,Dixon:2022rse}.) The duality involves a mathematically nontrivial \emph{antipode} map, which at the symbol level of polylogarithmic functions reverses the order of symbol letters, and roughly speaking, exchanges the roles of discontinuities and derivatives. The physical origin of such a duality is still not clear, thanks no less to the novelty of the antipode map in physically important examples. 

In this paper, we show that the antipode map appears in another context, that it is part of a new hidden symmetry of two-loop MHV amplitudes in planar $\cN=4$ SYM when evaluated on parity-even kinematics. We explicitly find the symmetry map for amplitudes up to eight external particles at the symbol level, and conjecture its existence for all multiplicity. This is the first evidence for a role of the antipode map in full amplitudes beyond six particles. On the other hand, its fate beyond two loops is still not clear.

The paper is organized as follows. In Section~\ref{sec:review} we review the MHV amplitudes in planar $\cN=4$ SYM, and the class of functions they belong to, the multiple polylogarithms, or MPL. We briefly review the \emph{antipode} operation associated with the Hopf algebra structure of MPLs, especially its action on their \emph{symbols}. In Section~\ref{sec:main} we give a statement of the conjectured antipodal symmetry for all two-loop MHV amplitudes in parity-even kinematics, where the symmetry operator is given by the antipode map plus a letter map, which is just a change of basis of the symbol letters; we also comment on the uniqueness of the letter map. Then, in Section~\ref{sec:fan} we discuss the action of the letter map on the tropical geometry of the symbol letters. Interestingly, it acts as an isomorphism on their tropical fans. Next, in Section~\ref{sec:number} we present further evidence for the existence of the antipodal symmetry at all multiplicities. In particular, we count the number of allowed first and final entry letters in the symbol, and show that they are the same in parity-even kinematics. We conclude in Section~\ref{sec:discuss}, and discuss future directions.

We also provide three ancillary files, \texttt{letter\_map\_R62.txt}, \texttt{letter\_map\_R72.txt} and \texttt{letter\_map\_R82.txt}, which give the letter maps associated with the antipodal symmetry at six-, seven- and eight-points respectively.

\section{Review} \label{sec:review}
\subsection*{MHV Amplitudes in Planar $\cN=4$ SYM}
Amplitudes in the planar limit are color-ordered partial amplitudes, which are coefficients of $\text{Tr}(T^{a_1}T^{a_2}\cdots T^{a_n})$ in the color decomposition. The maximally helicity violating (MHV) amplitudes are those related by supersymmetric Ward identities to all-gluon amplitudes among which exactly two have negative helicities. At loop level, the amplitudes have infrared divergences, but the infrared divergent part is captured by the exponentiation of the one-loop amplitude \cite{Bern:2005iz}. In short, the $n$-point MHV amplitude takes the following factorized form,
\be
\cA_n^{\mathrm{MHV}} = \cA_n^{\mathrm{BDS}} \cdot e^{R_n} \,,
\ee
where $\cA_n^{\mathrm{BDS}}$, the BDS ansatz \cite{Bern:2005iz}, captures all infrared divergences; and $R_n$, the \emph{remainder function}, is divergence-free. Since the BDS ansatz is the exponentiation of the one-loop amplitude, $R_n$ starts at two loops; that is, 
\be
R_n = g^4 R_n^{(2)} + O(g^6) \,
\ee
where $g^2 \equiv {g_{\text{YM}}^{2}N_c}/{(16 \pi^{2})}$ is the planar coupling constant.

Due to the duality to Wilson loops, the remainder function is dual conformal invariant, and has no kinematic dependence for $n<6$. For $n\ge 6$, the kinematic degrees of freedom is the same as the number of independent conformal cross ratios, which is $3(n-5)$. This is reduced to $2(n-5)$ on the slice that preserves parity, see Appendix \ref{app:ope}. It is on this parity-even surface that the conjectured symmetry of this paper holds.

\subsection*{Multiple Polylogarithms and the Antipode Map}
The remainder function $R_n$ is believed to belong to a class of special functions called \emph{multiple polylogarithms} (MPL) \cite{Goncharov:2001iea}, based on analyses of the loop integrands \cite{Arkani-Hamed:2012zlh}. An MPL $F$ can be defined recursively by specifying its total derivative,
\be
\label{eq:def_mpl}
dF = \sum_{\phi} F^{\phi} d\ln\phi \,,
\ee
plus a base point of integration; the $F^{\phi}$ above are also MPLs. The \emph{symbol} of an MPL is also defined recursively,
\be
\cS(F) = \sum_{\phi} \cS(F^{\phi}) \otimes \phi \,,
\ee
where $F^{\phi}$ and $\phi$ are the same as in (\ref{eq:def_mpl}) above. By definition, the symbol of a constant is zero. Put another way, the symbol keeps all the information of an MPL in its definition (\ref{eq:def_mpl}), except for the base point at every step of integration. The $\phi$ that shows up in the entries of a symbol are called \emph{symbol letters}; collectively, all the symbol letters in the symbol are the \emph{symbol alphabet}. Since logarithms of products are additive, from (\ref{eq:def_mpl}) we have identities such as
\be \nonumber
\ldots\otimes(\phi_1\phi_2)\otimes\ldots =
\left(\ldots\otimes\phi_1\otimes\ldots\right) +
\left(\ldots\otimes\phi_2\otimes\ldots\right)
\ee
and
\be \nonumber
\ldots\otimes\frac{1}{\phi}\otimes\ldots =
- \left(\ldots\otimes\phi\otimes\ldots\right) \,.
\ee

MPLs are endowed with a Hopf algebra structure, which is a bialgebra together with an \emph{antipode} operation that satisfies certain axioms. In particular, the antipode map acts on each term of a symbol as,
\be \label{eq:antipode}
S(\phi_1\otimes\phi_2\otimes \ldots \otimes\phi_{k-1}\otimes\phi_k) =
(-1)^k \,
\phi_k\otimes\phi_{k-1}\otimes \ldots \otimes\phi_2\otimes\phi_1 \,,
\ee
and extends to the whole symbol by linearity.

\section{The Antipodal Symmetry} \label{sec:main}
The symbol of all two loop remainder functions $R_n^{(2)}$ are obtained in ref.~\cite{Caron-Huot:2011zgw},
\be
\cS(R_n^{(2)}) = r_n^{ijkl} \phi_i\otimes\phi_j\otimes\phi_k\otimes\phi_l \,,
\ee
where summation over repeated indices is assumed on the right-hand side, and the summation is over a symbol alphabet that depends on $n$. We will look at the remainder function in parity-even kinematics. At symbol level, this is equivalent to projecting out all parity-odd letters,
\be \label{eq:Rn2e}
\cS(R_{n,e}^{(2)}) = r_{n,e}^{ijkl} \phi_i\otimes\phi_j\otimes\phi_k\otimes\phi_l \,,
\ee
where the summation is now over a smaller alphabet consisting only of letters that are parity invariant.

Now we are ready to state the main conjecture of this paper: for any $n$, there exists a matrix $A_n^{ij}$ such that
\vspace{15pt}
\be \label{eq:main}
\cS(R_{n,e}^{(2)}) = \frac{1}{4} \, S\left(\cS(R_{n,e}^{(2)})\Big|_{\ln\phi_i \mapsto A_n^{ij} \ln\phi_j}\right) \,,
\vspace{15pt}
\ee
where $S$ is the antipode map (\ref{eq:antipode}). In terms of the tensor $r_{n,e}^{ijkl}$ in (\ref{eq:Rn2e}), this is,
%
\be \label{eq:main_tensor}
r_{n,e}^{ijkl} = \frac{1}{4} \, r_{n,e}^{l'k'j'i'} A_n^{i'i} A_n^{j'j} A_n^{k'k} A_n^{l'l} \,,
\ee
where again summation over repeated indices is assumed.
In words, this says that the two-loop remainder function in parity-even kinematics is invariant, up to an overall factor $1/4$, under a change of symbol letter basis plus the antipode map. The overall factor $1/4$ turns out to be hard to get rid of, as discussed below.
For the maps we have found through $n=8$, $A_n^{ij}$ satisfies,
\be \label{eq:mapSq}
A_n^{ij} A_n^{jk} = 2\cdot\delta^{ik} \,.
\ee
So if we apply the transformation on the right-hand side of (\ref{eq:main}) twice, we get,
\be
\cS(R_{n,e}^{(2)}) \mapsto \left(\frac{1}{4}\right)^2 \cdot 2^4 \cdot \cS(R_{n,e}^{(2)}) = \cS(R_{n,e}^{(2)}) \,,
\ee
where $2^4$ comes from (\ref{eq:mapSq}) and the fact that it acts on each of the four entries in the symbol (\ref{eq:Rn2e}). The antipodal symmetry is therefore a $\mathbb{Z}_2$ symmetry.

Let us note here that the map $\ln\phi_i \mapsto A_n^{ij} \ln\phi_j$ should not be thought of as a map of the underlying kinematics. For example, for $n=6$, the parity-even kinematics is parametrized by two variables $(T,S)$ (see Appendix~\ref{app:ope}). It is easy to see that there exists no change of variables $(T,S) \mapsto (f_1(T,S), f_2(T,S))$ that results in the letter map (\ref{eq:R62map}). The letter map therefore is nothing but a linear algebraic manipulation on the letter basis; this is seen more obviously in the tensorial form (\ref{eq:main_tensor}).

We have explicitly found the letter map $A_n^{ij}$ through $n=8$. The maps for $n=6$ and $n=7$ are given in Appendix~\ref{app:map}, and we also give all the maps through $n=8$ in ancillary files to this paper.

\subsection*{Uniqueness of the Letter Map}
The letter map $A_n$ that satisfies (\ref{eq:main}) is by no means unique. Since $R_n$ is invariant under dihedral transformations, $OA_nO'$ would also satisfy (\ref{eq:main}) for any $O, O'$ in the dihedral group $D_n$\footnote{To be technical, $O,O'$ are in the representation of the dihedral group on the space of symbol letters.}. However, for $n=6$, it is uniquely fixed up to a scaling, either by requiring that the letter map respects dihedral symmetry, or by requiring that it acts as an isomorphism on the normal fans of symbol letters, as described in Section~\ref{sec:fan}. If we further require the overall scaling to be a rational number, then there exists no scaling which sets the 1/4 in (\ref{eq:main}) to unity. It would be interesting to understand more about this overall factor.
For $n=7,8$, the letter map is uniquely fixed by requiring the normal fan isomorphism and OPE factorization. That is, the map should reduce to a lower-point case for letters that are missing some $\{T_i,S_i\}$ for a subset of $i$.

\section{Letter Map and Tropical Fans} \label{sec:fan}
First let us recall the definition of Newton polytopes. The Newton polytope of a polynomial
\be
f(x_1,x_2,\ldots,x_d) = \sum_i c_i x_1^{b_{i,1}} x_2^{b_{i,2}} \cdots x_d^{b_{i,d}}
\ee
is the convex hull of $\{\mathbf{b}_i\}$, where each $\mathbf{b}_i = (b_{i,1},b_{i,2},\ldots,b_{i,d})$ is a point in $d$-dimensional space. The \emph{normal fan} of a polytope consists of a set of \emph{rays}, where each ray is a $d$-dimensional vector (modulo positive scaling), and a set of \emph{cones}, where each cone is a subset of the rays. The cones are in bijection with the faces of the polytope. For example, each facet (i.e. a $(d-1)$-dimensional face of the polytope) corresponds to a cone consisting of a single ray that is the inner normal vector of this facet.\footnote{Convex hulls and normal fans can be conveniently computed using computer programs such as \texttt{polymake} \cite{polymake:2000}.}

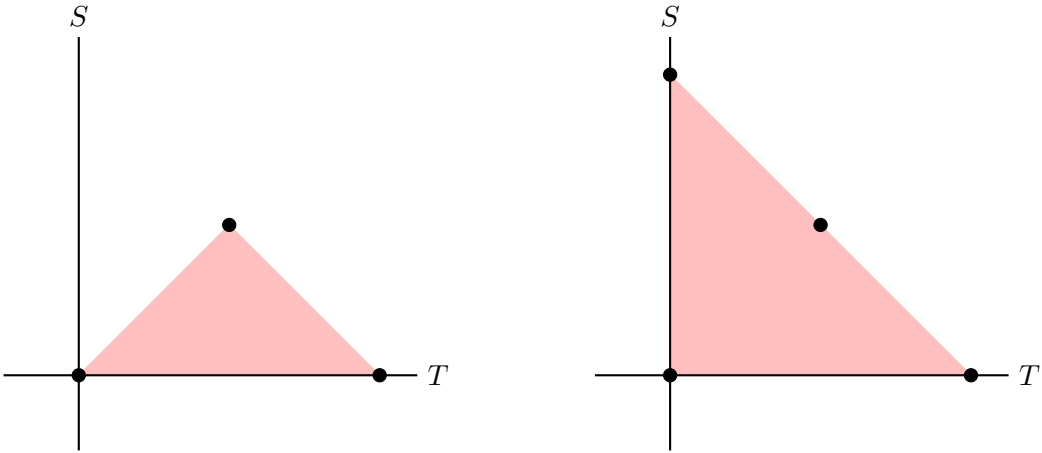
\begin{figure}[ht]
\begin{subfigure}{.5\textwidth}
\centering
\begin{tikzpicture}
\filldraw[pink] (0,0) -- (4,0) -- (2,2) -- cycle;
\filldraw[black] (0,0) circle (2.5pt);
\filldraw[black] (4,0) circle (2.5pt);
\filldraw[black] (2,2) circle (2.5pt);
\draw[thick] (-1,0) -- (4.5,0) node[anchor=west]{$T$};
\draw[thick] (0,-1) -- (0,4.5) node[anchor=south]{$S$};
\end{tikzpicture}
\end{subfigure}
\begin{subfigure}{.5\textwidth}
\centering
\begin{tikzpicture}
\filldraw[pink] (0,0) -- (4,0) -- (0,4) -- cycle;
\filldraw[black] (0,0) circle (2.5pt);
\filldraw[black] (4,0) circle (2.5pt);
\filldraw[black] (0,4) circle (2.5pt);
\filldraw[black] (2,2) circle (2.5pt);
\draw[thick] (-1,0) -- (4.5,0) node[anchor=west]{$T$};
\draw[thick] (0,-1) -- (0,4.5) node[anchor=south]{$S$};
\end{tikzpicture}
\end{subfigure}
\caption{The Newton polytopes of the polynomials $1+ST+T^2$ (left) and $1+S^2+2ST+T^2$ (right). Each black dot corresponds to a monomial, and the pink shaded area is their convex hull.}
\label{fig:newton}
\end{figure}

Let us look at some examples. $1+ST+T^2$ is a polynomial that shows up as a six-point symbol letter in (\ref{eq:R62letters}). The three monomials $1$, $ST$, $T^2$ correspond to three points $(0,0)$, $(1,1)$, $(2,0)$ respectively in the $T$-$S$ plane, as can be seen as the black dots in the left panel of Figure~\ref{fig:newton}, and the Newton polytope is the pink shaded area in the figure. Similarly, the right panel of Figure~\ref{fig:newton} shows the Newton polytope of $1+S^2+2ST+T^2$, which is also a symbol letter at six points.

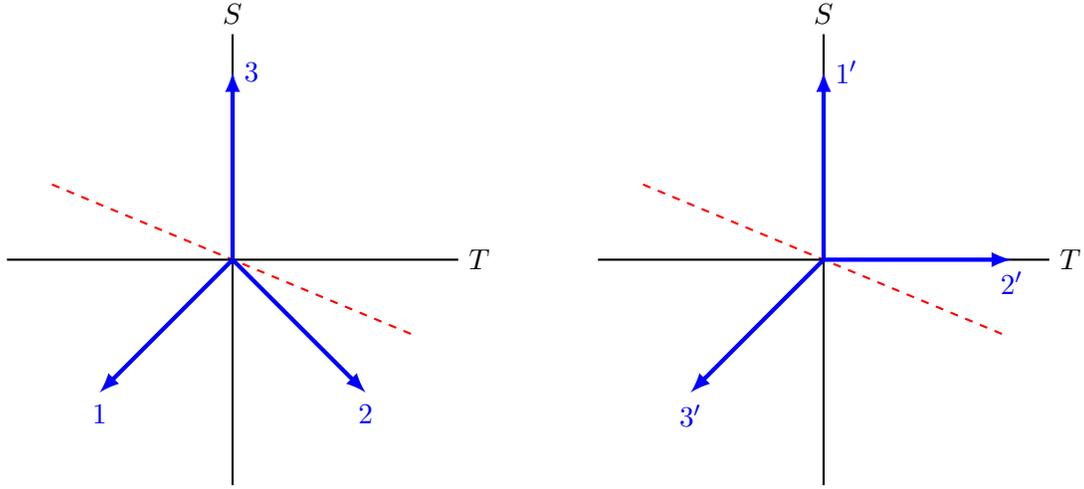
\begin{figure}[ht]
\begin{subfigure}{.5\textwidth}
\centering
\begin{tikzpicture}
\draw[red, thick, dashed] (-2.4,1) -- (2.4,-1);
\draw[thick] (-3,0) -- (3,0) node[anchor=west]{$T$};
\draw[thick] (0,-3) -- (0,3) node[anchor=south]{$S$};
\draw[ultra thick, blue, -latex] (0,0) -- (-1.77,-1.77) node[anchor=north]{1};
\draw[ultra thick, blue, -latex] (0,0) -- (1.77,-1.77) node[anchor=north]{2};
\draw[ultra thick, blue, -latex] (0,0) -- (0,2.5) node[anchor=west]{3};
\end{tikzpicture}
\end{subfigure}
\begin{subfigure}{.5\textwidth}
\centering
\begin{tikzpicture}
\draw[red, thick, dashed] (-2.4,1) -- (2.4,-1);
\draw[thick] (-3,0) -- (3,0) node[anchor=west]{$T$};
\draw[thick] (0,-3) -- (0,3) node[anchor=south]{$S$};
\draw[ultra thick, blue, -latex] (0,0) -- (0,2.5) node[anchor=west]{$1'$};
\draw[ultra thick, blue, -latex] (0,0) -- (2.5,0) node[anchor=north]{$2'$};
\draw[ultra thick, blue, -latex] (0,0) -- (-1.77,-1.77) node[anchor=north]{$3'$};
\end{tikzpicture}
\end{subfigure}
\caption{The normal fans of the Newton polytopes of $1+ST+T^2$ (left) and $1+S^2+2ST+T^2$ (right). The rays are drawn in blue arrows; they are the inner normal vectors of the facets of Newton polytopes in Figure~\ref{fig:newton}. The cones are $\{\{1\},\{2\},\{3\},\{1,2\},\{2,3\},\{3,1\}\}$ and $\{\{1'\},\{2'\},\{3'\},\{1',2'\},\{2',3'\},\{3',1'\}\}$ respectively. The fan on the left is isomorphic to the fan on the right by the \textbf{flip} along the red dashed line, the line at an angle $-\pi/8$ with the $T$-axis.}
\label{fig:fan}
\end{figure}

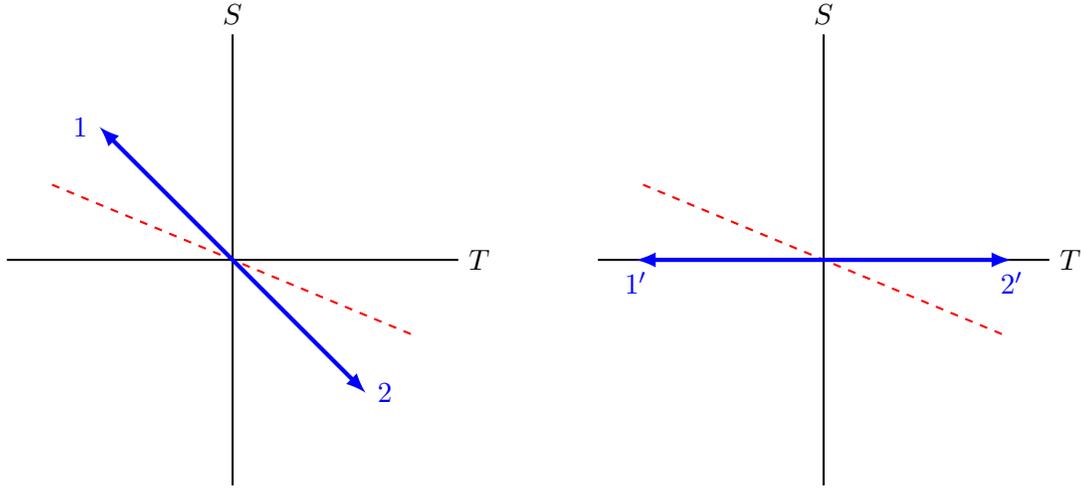
\begin{figure}[ht]
\begin{subfigure}{.5\textwidth}
\centering
\begin{tikzpicture}
\draw[red, thick, dashed] (-2.4,1) -- (2.4,-1);
\draw[thick] (-3,0) -- (3,0) node[anchor=west]{$T$};
\draw[thick] (0,-3) -- (0,3) node[anchor=south]{$S$};
\draw[ultra thick, blue, -latex] (0,0) -- (-1.77,1.77) node[anchor=east]{1};
\draw[ultra thick, blue, -latex] (0,0) -- (1.77,-1.77) node[anchor=west]{2};
\end{tikzpicture}
\end{subfigure}
\begin{subfigure}{.5\textwidth}
\centering
\begin{tikzpicture}
\draw[red, thick, dashed] (-2.4,1) -- (2.4,-1);
\draw[thick] (-3,0) -- (3,0) node[anchor=west]{$T$};
\draw[thick] (0,-3) -- (0,3) node[anchor=south]{$S$};
\draw[ultra thick, blue, -latex] (0,0) -- (-2.5,0) node[anchor=north]{$1'$};
\draw[ultra thick, blue, -latex] (0,0) -- (2.5,0) node[anchor=north]{$2'$};
\end{tikzpicture}
\end{subfigure}
\caption{The normal fans of the Newton polytopes $S+T$ (left) and $1+T^2$ (right). The rays are drawn in blue arrows; the cones are $\{\{1\},\{2\}\}$ and $\{\{1'\},\{2'\}\}$ respectively. The two fans are isomorphic to each other, by the \textbf{flip} along the same red dashed line as in Figure~\ref{fig:fan}.}
\label{fig:fan2}
\end{figure}

The normal fans of the Newton polytopes of $1+ST+T^2$ and $1+S^2+2ST+T^2$ are shown in Figure~\ref{fig:fan}. The two fans are actually isomorphic to each other, under the \textbf{flip} along the red dashed line shown in the figure. Interestingly, the letter map (\ref{eq:R62map}) generates exactly this isomorphism; that is, the normal fans of the letters $1+ST+T^2$ and $1+S^2+2ST+T^2$ are mapped into each other under the letter map\footnote{On the right-hand side of (\ref{eq:R62map}), the normal fan of $\frac{\left(1+S T+T^2\right)^2}{S^2 T^2}$ is just that of $1+S T+T^2$. To see this, note that the overall square in the numerator of $\frac{\left(1+S T+T^2\right)^2}{S^2 T^2}$ simply rescales the Newton polytope of $1+S T+T^2$, and the $T$ and $S$ in the denominator is simply a translation; scaling and translation have no effect on the normal fan of the polytope. Similarly, the normal fan of $\frac{1+S^2+2 S T+T^2}{S^2}$ is the same as $1+S^2+2 S T+T^2$.}. Furthermore, the letter map generates an isomorphism between the normal fans of $S+T$ and $1+T^2$ as well, and the isomorphism is again given by the same \textbf{flip} above, see Figure~\ref{fig:fan2}. Including the trivial cases of $T$ and $S$, the letter map acts as an isomorphism on the normal fans of the symbol letters; and the isomorphism is always given by the \textbf{flip}.

This observation generalizes to higher multiplicities (at least through eight points where we have explicit letter maps to check against) that the letter map acts as an isomorphism on the normal fans of letters, with the isomorphism given by,
%
\be
\textbf{flip}_1 \otimes \textbf{flip}_2 \otimes \cdots \otimes \textbf{flip}_{n-5} \,,
\ee
where $\textbf{flip}_i$ is the same \textbf{flip} as in the six-point case but in the $T_i$-$S_i$ plane.

\section{Evidence at All Multiplicities} \label{sec:number}
Since the equality (\ref{eq:main}) involves an antipode map, the order of the symbol entries are reversed on the right-hand side; the first entries of the symbol become the last entries, and vice versa. The symbol letters that can appear in first entries and last entries are heavily constrained by physics. These are often called the \emph{first-entry conditions} and the \emph{final-entry conditions} in the literature. For Eq.~(\ref{eq:main}) to hold, the number of first entries and the number of final entries must be the same. As we will see in this section, this is indeed the case in parity-even kinematics.

The first entries of a symbol encode its branch points. For a physical amplitude, its branch points correspond to vanishing Mandelstam variables,
\be \label{eq:num_first}
s_{i,i+1,\ldots,j} = (p_i+p_{i+1}+\cdots+p_j)^2 \propto \left<i-1,i,j-1,j\right> \,.
\ee
where $\left<i,j,k,l\right> \equiv \det(Z_i Z_j Z_k Z_l)$ and $Z_1,\ldots,Z_n$ are momentum twistors \cite{Hodges:2009hk}; all first entries are therefore of the form $\left<i-1,i,j-1,j\right>$. These letters come in two different cases. The first case is of the form $\left<i-1,i,i+1,i+2\right>$; there are $n$ such letters for $i=1,\ldots,n$. The second case is when $(i-1,i)$ and $(j-1,j)$ are \emph{not} adjacent to each other, e.g. $\left<i-1,i,j-1,j\right>=\left<1,2,4,5\right>$; there are $n(n-5)/2$ such letters.\footnote{The 2 in the denominator comes from the symmetry $\left<i-1,i,j-1,j\right> = \left<j-1,j,i-1,i\right>$.} Since the remainder function is conformal invariant, only conformal invariant combinations of such letters can show up, i.e. the little group weight of every $Z_i$ has to drop out; this take $n$ out of the total number of allowed letters. Overall, the number of allowed first entries is,
\be
\text{\# of first entries } = n + \frac{n(n-5)}{2} - n = \frac{n(n-5)}{2} \,.
\ee

On the other hand, the allowed final entries are constrained to be of the form $\left<i-1,i,i+1,j\right>$ \cite{Caron-Huot:2011dec}. Let us count the number of allowed letters. First, again there are $n$ letters of the form $\left<i-1,i,i+1,i+2\right>$. Next, when $(i-1,i,i+1)$ are not next to $j$, there are $n(n-5)$ such letters. And again we have to take $n$ out of these for conformal invariance. Overall,
\be
\text{\# of final entries } = n + n(n-5) - n = n(n-5) \,.
\ee
Now, the parity operation preserves letters of the form $\left<i-1,i,j-1,j\right>$, and maps \cite{Golden:2013xva}
\be
\left<i-1,i,i+1,j\right> \leftrightarrow \left<i,j-1,j,j+1\right> \,.
\ee
So the number of parity-even final entries is,
\be
\text{\# of parity-even final entries } = n + \frac{n(n-5)}{2} - n = \frac{n(n-5)}{2} \,,
\ee
which is exactly the same as the number of parity-even first entries. This therefore supports a symmetry of $R_n$ in parity-even kinematics that involves the antipode map, such as (\ref{eq:main}).

\section{Discussion} \label{sec:discuss}
In this paper we conjecture an antipodal symmetry (\ref{eq:main}) for all two-loop remainder functions in parity-even kinematics. We explicitly check the existence of such a symmetry at symbol level through eight points, and provide some all-multiplicity evidence.

First, let us emphasize the presence of the antipode map in this symmetry. The antipode map arises naturally as part of the Hopf algebra structure of multiple polylogarithms, and has a long history in the study of such functions in the mathematical literature. However, its importance in physics has been under-explored. One of the rare examples where the antipode map shows up is in the ``antipodal duality'' of ref.~\cite{Dixon:2021tdw}. In this paper we give another such example. Given the ubiquity of MPLs in physical quantities, we should expect the antipode map to play a much bigger role than we have currently understood. It is also interesting that our symmetry map at six points (\ref{eq:R62mapOld}) bears resemblance to the map in six-particle amplitude/three-particle form factor duality \cite{Dixon:2021tdw}. One might wonder whether the antipodal symmetry for higher-point amplitudes could potentially hint at a general higher-point ampitude/form factor duality.

Further, the antipodal symmetry seems to point to some deep structures of the amplitudes, especially in connection with tropical geometry. Indeed, it has been known for a while that some information of the amplitudes is encoded in certain tropical polytopes (or their normal fans) \cite{Arkani-Hamed:2019rds,Drummond:2019cxm,Henke:2019hve}. These are polytopes that are associated with the amplitude as a whole; specifically, they are Minkowski sums of the Newton polytopes of subsets of symbol letters. However, the letter map in (\ref{eq:main}) preserves the tropical fan structures \emph{letter by letter}. This suggests there might be information contained in the tropical geometry of individual symbol letters, that has yet to be studied.

Many things are still unsatisfactory. For example, the symmetry involves a non-trivial letter map, which is a change of basis of the symbol alphabet, but the map cannot be obtained from a change of variables of the underlying kinematics; this makes it difficult to lift the symmetry to the full function level. It might be possible to shed some light on this problem by looking at the full functional form of some two-loop remainder functions \cite{Goncharov:2010jf,Golden:2013xva,Golden:2018gtk,Golden:2021ggj} and the cluster algebraic structures associated with them, which had played an important role in either simplifying or constructing these full functions. Also, the naive generalization of (\ref{eq:main}) does not seem to work beyond two loops, starting at three loops and six points. It would be interesting to see if there exists a clever normalization of amplitudes\footnote{One guess would be the BDS-like normalization \cite{Alday:2009dv,Yang:2010as}. The antipodal symmetry holds for one- and two-loop BDS-like-normalized amplitudes at six and seven points, but still fails at three loops. Also, the BDS-like normalization is not defined for $n$ a multiple of four (see also \cite{Golden:2019kks}).} (or maybe a more complicated transformation) that generalizes the antipodal symmetry to an all-loop statement.

\acknowledgments
I would like to thank Lance Dixon for stimulating discussions and comments on the draft, and Lance Dixon, {\"O}mer G{\"u}rdo{\u{g}}an, Andrew McLeod and Matthias Wilhelm for collaboration on related projects. This work was supported by the US Department of Energy under contract DE--AC02--76SF00515, and in part by the National Science Foundation under Grant No. PHY-1748958. I am also grateful to the Kavli Institute for Theoretical Physics, TASI 2022 and the University of Colorado Boulder for hospitality.

\appendix
\section{OPE Variables} \label{app:ope}
It is helpful to parametrize the $n$-particle scattering kinematics in terms of Pentagon OPE variables. Recall that the $n$-particle MHV amplitude is dual to an $n$-gon Wilson loop with null edges, which can be thought of as a sum over flux tube states living on the middle $n-5$ null squares plus transitions (Pentagon transitions) among them \cite{Basso:2013vsa}. The overall kinematics is then given by those of the middle squares. Each square is characterized by three parameters, $(\tau_i,\sigma_i,\phi_i)$, which are the conformal time and space, and the rotation angle in the transverse plane respectively. There is then a total of $3(n-5)$ kinematic degrees of freedom. Using these, we define,
\be
T_i = e^{-\tau_i},\ 
S_i = e^{\sigma_i},\ 
F_i = e^{i\phi_i}  \qquad \text{for $i = 1,\ldots,n-5$} \,.
\ee
All symbol letters at two loops are rational functions of $\{T_i,S_i,F_i\}$. Furthermore, $T_i$ and $S_i$ are parity-even and $F_i$ is parity-odd (that is, $F_i\to 1/F_i$ under parity). To go to parity-even kinematics, in OPE variables we simply set all $F_i\to1$, and look at the reduced symbol alphabet. The parity-even surface has $2(n-5)$ kinematic degrees of freedom parametrized by $\{T_i,S_i\}$.

\section{Letter Maps for $R_{6,e}^{(2)}$ and $R_{7,e}^{(2)}$} \label{app:map}
The symbol alphabet for $R_6$ is usually given in the following basis (see e.g. ref.~\cite{Caron-Huot:2019vjl}),
\be \label{eq:R62lettersOld}
\{u,\,v,\,w,\,1-u,\,1-v,\,1-w,\,y_u,\,y_v,\,y_w\} \,,
\ee
where $(u,v,w)$ are the three conformal cross ratios. The $y_i$ are parity-odd and become 1 in parity-even kinematics. After setting $F$ to 1, the parity-even letters are related to $T$ and $S$ by,
\begin{align}
u&= \frac{1}{1+S^2+2 S T+T^2}, &
1-u&= \frac{(S+T)^2}{1+S^2+2S T+T^2},\nonumber\\
v&= \frac{S^2}{\left(1+T^2\right) \left(1+S^2+2 S T+T^2\right)}, &
1-v&= \frac{\left(1+S T+T^2\right)^2}{\left(1+T^2\right) \left(1+S^2+2 S T+T^2\right)}, \nonumber\\
w&= \frac{T^2}{1+T^2}, &
1-w&= \frac{1}{1+T^2} \,.
\end{align}
We can take the following symbol alphabet basis,
\be \label{eq:R62letters}
\{S,\ T,\ S+T,\ 1+T^2,\ 1+S T+T^2,\ 1+S^2+2 S T+T^2\} \,.
\ee
In this basis, the matrix $A_6^{ij}$ that satisfies (\ref{eq:main}) is
\be
A_6^{ij} =
\begin{pmatrix}
-1 & -1 & 0 & 0 & 0 & 0 \\
-1 &  1 & 0 & 0 & 0 & 0 \\
-1 & -1 & 0 & 1 & 0 & 0 \\
-2 &  0 & 2 & 0 & 0 & 0 \\
-2 &  0 & 0 & 0 & 0 & 1 \\
-2 & -2 & 0 & 0 & 2 & 0
\end{pmatrix} \,.
\ee
Or to write out $\ln\phi_i \mapsto A_6^{ij} \ln\phi_j$ explicitly (and getting rid of the logarithms), the letter map is,
\begin{align} \label{eq:R62map}
&S \mapsto \frac{1}{S T}, &
&T \mapsto \frac{T}{S}, \nonumber\\
&S+T \mapsto \frac{1+T^2}{S T}, &
&1+T^2 \mapsto \frac{(S+T)^2}{S^2}, \nonumber\\
&1+S T+T^2 \mapsto \frac{1+S^2+2 S T+T^2}{S^2}, &
&1+S^2+2 S T+T^2 \mapsto \frac{\left(1+S T+T^2\right)^2}{S^2 T^2} \,.
\end{align}
After translating to the more conventional basis (\ref{eq:R62lettersOld}) and applying a flip transformation, the letter map (\ref{eq:R62map}) can be written as,
\be \label{eq:R62mapOld}
u \mapsto \frac{v w}{(1-v)(1-w)} \,, \quad
1-u \mapsto \frac{u}{(1-v)(1-w)} \,,
\ee
plus cyclic images. This looks exactly like the duality map between the six-particle amplitude and the three-particle form factor in ref.~\cite{Dixon:2021tdw}, except that now both sides of the map are interpreted as letters on the amplitude side.

For $R_7$, the symbol alphabet can be written in a basis of 42 letters $g_{ij}$ \cite{Dixon:2020cnr}, where each $\{g_{ij},j=1,\ldots,7\}$ is in a cyclic orbit of $D_7$. $g_{5j}$ and $g_{6j}$ are parity odd, and the rest are related to cross ratios by,
\begin{align}
g_{1j} &= u_j\,, &
g_{3j} &= 1 - u_{j+2}u_{j+5}\,,  \nonumber\\
g_{2j} &= 1 - u_{j}\,, &
g_{4j} &= 1 - u_{j+1}u_{j+4} - u_{j+3}u_{j+6}\,. 
\end{align}
In this basis, the letter map $\ln\phi_i \mapsto A_7^{ij} \ln\phi_j$ is,
\begin{align}
g_{11} &\mapsto \frac{g_{14}g_{15}g_{21}}{g_{24}g_{25}}\,, &
g_{31} &\mapsto \frac{g_{11}g_{41}}{g_{22}g_{27}}\,,  \nonumber\\
g_{21} &\mapsto \frac{g_{13}g_{16}}{g_{24}g_{25}}\,, &
g_{41} &\mapsto \frac{g_{12}g_{17}g_{31}^2}{g_{21}g_{23}g_{26}}\,. 
\end{align}
plus cyclic images. The letter map in terms of OPE variables is given in an ancillary file to this paper.

\bibliographystyle{JHEP}
\bibliography{../generate_bib/all}

\end{document}